\documentclass{aa}
\usepackage{graphicx}
\usepackage{natbib}
\bibpunct{(}{)}{;}{a}{}{,}
\begin{document}
\title{Radio spectral properties and the magnetic field of the SNR S147\thanks{Figs.~1 and 2 are available
in FITS format at the CDS via anonymous ftp to cdsarc.u-strasbg.fr (130.79.128.5) or via http://cdsweb.u-strasbg.fr/cgi-bin/qcat?J/A+A/ }}
\author{L. Xiao\inst{1}
        \and E. F\"urst\inst{2}
    \and W. Reich\inst{2}
    \and J. L. Han\inst{1}}
\titlerunning{Spectral properties and the magnetic field of the SNR S147}

\institute{ National Astronomical Observatories, Chinese Academy of
  Sciences, Jia-20 DaTun Road, Chaoyang District, Beijing 100012, China.
            \email{xl,hjl@bao.ac.cn}
            \and Max-Planck-Institut f\"{u}r Radioastronomie,
             Auf dem H\"ugel 69, 53121 Bonn, Germany \\
            \email{efuerst,wreich@mpifr-bonn.mpg.de}
          }

\date{Received / Accepted}
\offprints{W. Reich}

\abstract
{S147 is a large faint shell-type supernova remnant (SNR) known for its remarkable
spectral break at cm-wavelength, which is an important physical property to
characterize SNR evolution. However, the spectral break is based on
radio observations with limited precision.}
{New sensitive observations at high frequencies are required for a detailed
study of the spectral properties and the magnetic field structure of S147.}
{We conducted new radio continuum and polarization observations of S147 at
$\lambda$11\ cm and at $\lambda$6\ cm with the Effelsberg 100-m telescope
and the Urumqi 25-m telescope, respectively. We combined these new data
with published lower-frequency data from the Effelsberg 100-m telescope,
and with very high-frequency data from WMAP to investigate the spectral turnover
and polarization properties of S147.
}
{S147 consists of numerous filaments embedded in diffuse emission. We found
that the integrated flux densities of S147 are $34.8\pm 4.0~$Jy at
$\lambda$11\ cm and $15.4\pm 3.0~$Jy at $\lambda$6\ cm. These new measurements
confirm the known spectral turnover at $\sim$1.5\ GHz, which can be entirely
attributed to the diffuse emission component. The spectral index above the turnover is
$\alpha = -1.35 \pm 0.20~(\rm S\sim\nu^{\alpha})$.  The filamentary
emission component has a constant spectral index over the entire
wavelength range up to 40.7~GHz of $\alpha = -0.35 \pm 0.15$. 
The weak polarized emission of S147 is at the same level as the ambient
diffuse Galactic polarization. The rotation measure of the eastern 
filamentary shell is about $-70$\ rad m$^{-2}$.  }
{The filamentary and diffuse emission components of S147 have
different physical properties, which make S147 outstanding among shell type SNRs. 
We attribute the weak polarization of S147 at $\lambda$11\
cm and at $\lambda$6\ cm to a section of the S147 shell showing 
a tangetial magnetic field direction.
}


\maketitle

\section{Introduction}

The extended object Shajn 147 (S147) has long, beautiful delicate filaments
visible in the optical bands, and is located in the anti-center of the
Galaxy. It was first classified as a possible shell-type supernova remnant
(SNR) by \citet{m+58} and \citet{v+60}. The distance to S147 was estimated
to be around 1~kpc \citep{cc76,k+80}, using the so called $\Sigma$-D relation
\citep{milne+79}. The interstellar reddening toward the SNR \citep{f+85}
suggests a smaller distance of 0.8~kpc. This is supported by absorption lines
of the B1e star HD36665 at 880~pc originating from high velocity gas of S147 
\citep{p+81,sw+04}. The expansion velocity of the shell is about 80 - 120 km
s$^{-1}$ \citep{p+81,ka+79}. The age of S147 was estimated to be $\sim 2 \times
10^5$~yr through the radiative blast wave's Sedov-Taylor solution
\citep{sf+80,k+80}. The PSR J0538+2817 has been discovered within the boundary
of S147 \citep{a+96}, and is probably physically associated with the SNR
\citep{a+96,n+07}. The kinematic age of the pulsar, if originated from the
SNR center, is about $\sim 4\times 10^4$~yr. The parallax of the pulsar
suggests a distance of $1.47_{-0.27}^{+0.42}$~kpc \citep{n+07}.

\citet{d+74} summarized early radio maps of S147 between 178~MHz and 1420~MHz. 
She obtained a flux density spectral index of $\alpha= -0.67\pm 0.37$, revealing 
S147 as a SNR. Later, radio observations of S147 at 1648\ MHz and 2700\ MHz
were made by \citet{k+80}, and the southern part of the SNR was observed at
4995\ MHz by \citet{sf+80}.  These high-frequency observations suggested a
spectral turnover near 1.5\ GHz with a flat spectrum at lower frequencies and
a steep spectrum at higher frequencies. The turnover needs to be confirmed
due to large uncertainties in the low-frequency measurements and the
fact that the observation at 4995\ MHz covered only  the southern part of the remnant.
Later sensitive measurements, at 1425\ MHz and 2695\ MHz, by \citet{fr+86}
could not improve the integrated radio spectrum, but revealed spectral index
variations across S147. Regions of bright filaments show a flux density
spectral index of about $\alpha = -0.5$, while regions in between show a
spectral index near $\alpha = -1.0$. The observation at 863\ MHz by
\citet{rzf+03} provided a more precise flux density at a low frequency, but did not
improve on the uncertainty in the spectrum. The 2695\ MHz radio map made by
\citet{fr+86} for the first time showed linearly polarized emission associated with some
filamentary structure. Detailed information from this
polarization observation cannot be derived due to an insufficient signal-to-noise
ratio.

One goal of this work is to use the new sensitive $\lambda$6\ cm receiver at
the Urumqi 25-m radio telescope to obtain a complete radio map at this band
for a detailed study of the spectral turnover. Here we present results of
total power and linear polarization maps at 4800\ MHz ($\lambda$6\ cm). We also present 
new observations at 2639\ MHz ($\lambda$11\ cm) with the Effelsberg 100-m radio telescope and
unpublished polarization data at 1420\ MHz ($\lambda$21\ cm)  obtained from the ``Effelsberg
Medium Galactic Latitude Survey" \citep{rf+04}. The new observations and the
data processing are described in Sect.~2. The investigation of the
integrated radio spectrum and possible spectral variations across the
remnant are discussed in Sect.~3. The radio polarization and the rotation
measure toward S147 are analyzed in Sect.~4. The results are summarized in
Sect.~5.

\section{Observations and data processing}

New observations of S147 have been made with the Urumqi 25-m radio telescope
at $\lambda$6\ cm and Effelsberg 100-m radio telescope at $\lambda$11\ cm. The main
parameters are listed in Table~\ref{obspara}.
\begin{table}
   \caption{Observational parameters}
   \label{obspara}
  {\begin{tabular}{lll}\hline\hline
  Wavelength                      & $\lambda$6\ cm & $\lambda$11\ cm         \\
  Frequency                       & 4800\ MHz    & 2639\ MHz       \\
  Bandwidth                       & 600\ MHz     & 80\ MHz         \\
  HPBW [\arcmin ]                 & 9.5          & 4.4             \\
  aperture efficiency[\%]        & 62           &  53             \\
  beam efficiency[\%]             & 67           &  58             \\
  T$_{\rm sys}$[K]                & 22           &  17             \\
  T$_{B}[K]$/S[Jy]                & 0.164        & 2.60            \\
  Main Calibrator                 & 3C286        & 3C286           \\
  \ \ \   Flux Density            & 7.5\ Jy      & 11.5\ Jy        \\
  \ \ \   Polarization Percentage & 11.3\%       & 9.9\%           \\
  \ \ \   Polarization Angle      & 33$\degr$    & 33$\degr$       \\
  No. of coverages                &  14          & 6               \\
  pixel integration time [sec]     & 18.6         & 3               \\
  r.m.s. (total intensity) [mK]                & 0.7$^*$          &  5.0            \\
  r.m.s. (polarized intensity)  [mK]                & 0.3$^*$          &  3.5            \\\hline
\multicolumn{3}{l}{$^*$The quoted sensitivity at $\lambda$6\ cm is valid for the area} \\
\multicolumn{3}{l}{l=$178{\fdg}0$ to
l=$182{\fdg}2$, b=-$3{\fdg}6$ to b=-$0{\fdg}6$ (see text)} \\
  \end{tabular}}
\end{table}

\subsection{Observations at $\lambda$6\ cm}

We made continuum and polarization observations of S147 between January 2005 and February 2006.
We scanned the large area of S147 along Galactic longitude or
Galactic latitude direction. Five coverages of the field of ($\Delta l \times
\Delta b$) = ($5\degr \times 5\degr$) centered on ($l,b$) = $180{\fdg}2
,-1{\fdg}7$ were observed in January 2005, four coverages ($\Delta l\times
\Delta b$) = ($4{\fdg}2 \times 4{\fdg}2$) centered on ($l,b$) = $180{\fdg}1
,-1{\fdg}5$ were added at the end of 2005, and five additional maps of
($\Delta l\times \Delta b$) = ($5\degr \times 5\degr$) centered on ($l,b$) =
$180{\fdg}2,-1{\fdg}6$ were observed again in February 2006. We moved the telescope
with a scanning velocity between $2\arcmin$/s and
$3\arcmin$/s. The sampling and the pixel size of the map were
$3\arcmin$. There is one map scanned with a narrow bandwidth (295\ MHz
instead of 600\ MHz) to mask interference from InSat \citep[see][]{s+07}, so
the effective total integration time is about 18.6~s per pixel and the
theoretical sensitivity is about 0.2\ mK~T$_{\rm A}$, using $\sigma_I=T_{\rm
sys}/\sqrt{\Delta\nu\tau}$, where $T_{\rm sys}$ is the system temperature,
T$_{\rm A}$ is the antenna temperature, $\Delta\nu$ is the bandwidth, and
$\tau$ is the integration time. The quoted sensitivity corresponds to 0.3\
mK~T$_{\rm B}$, where T$_{\rm B}$ is the brightness temperature.

The raw data from the four backend channels (RR$^*$, LL$^*$, RL$^*$,
LR$^*$), together with time information and telescope position were stored
in the MBFITS format \citep{mph+07}, and subsequently converted into the NOD2 format
\citep{h+74} to be further processed by the NOD2 based data reduction
package under a Linux operation system. We edited the individual maps to
remove any interference, to correct baseline curvatures by polynomial
fitting and to suppress ``scanning effects" by applying an ``unsharp masking
method" \citep{sr+79} especially for the U and Q maps. Finally, we re-tabulated all maps
into a map size of $5\degr \times5\degr $ centered on
$l,b=180{\fdg}2,-1{\fdg}7$ with a pixel size of 4\arcmin, and ``weaved" all maps 
together applying the method of \citet{eg+88}, which ``destripes" a set of maps
in the Fourier domain. We finally corrected the polarization map from the
positive noise bias following \citet{wk+74}.

\begin{figure}
\centering
\includegraphics[width=8cm]{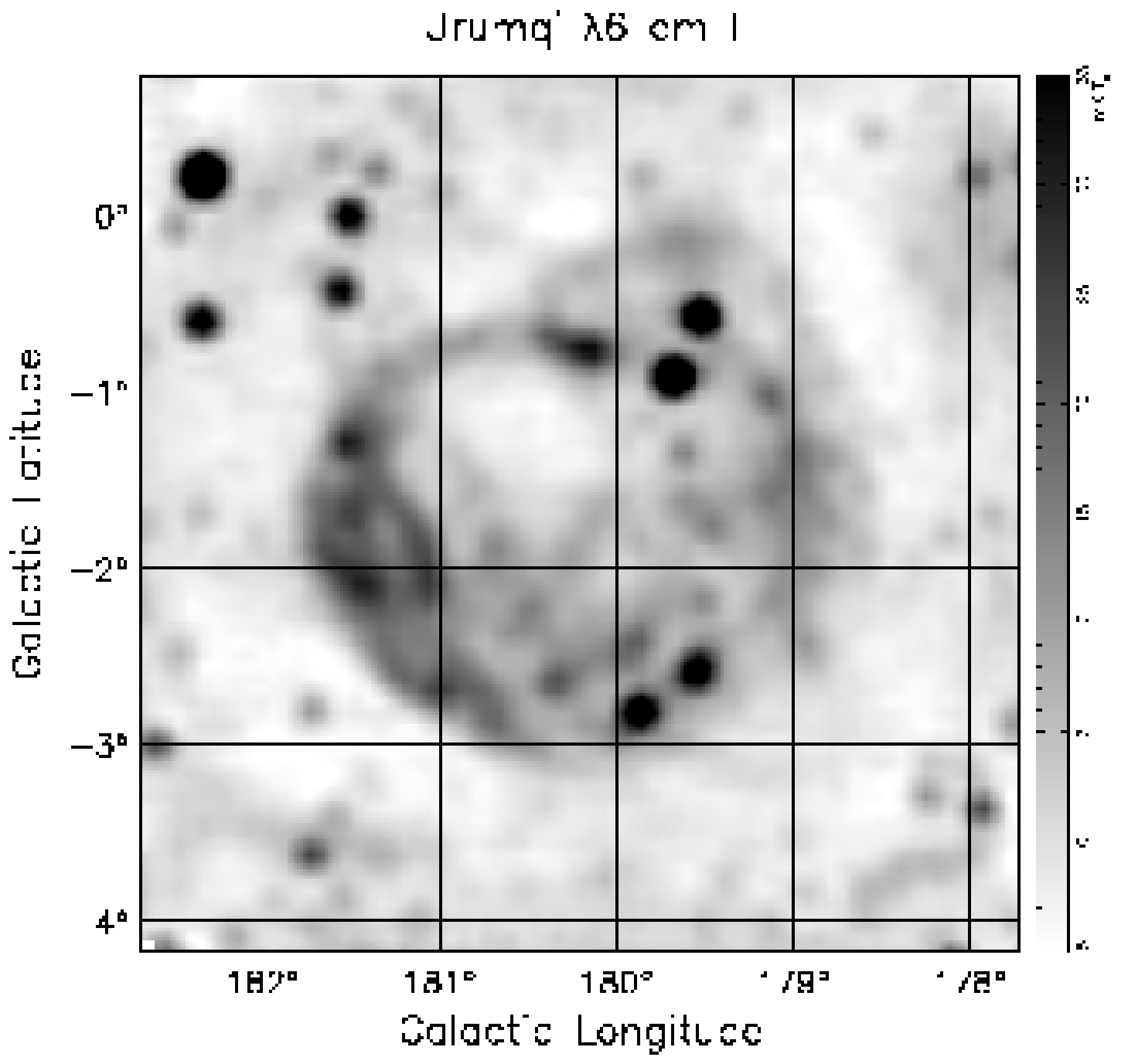}
\includegraphics[width=8cm]{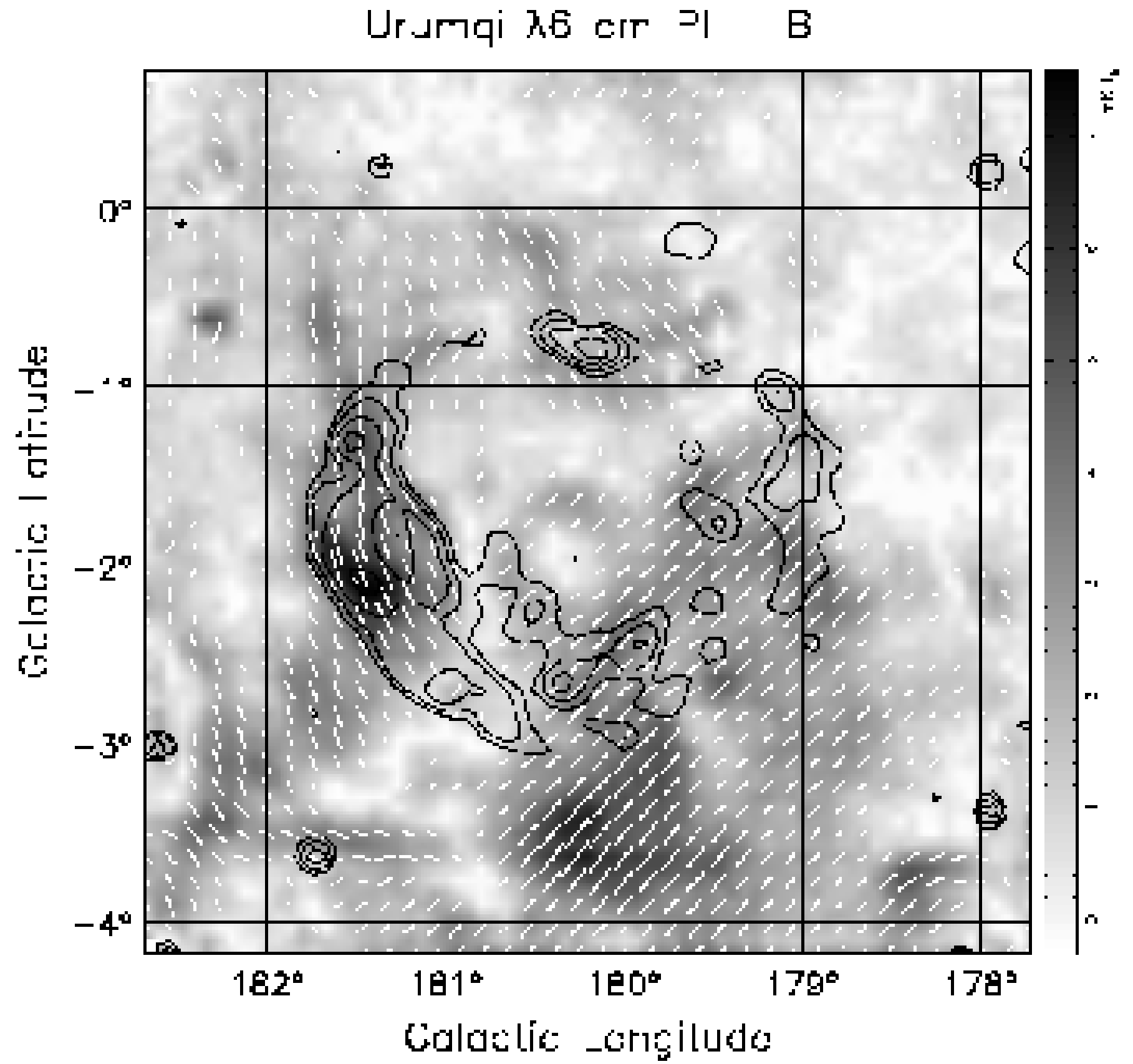}
    \caption{The Urumqi $\lambda$6\ cm map of S147 at an angular resolution of
    9\farcm5. The upper panel shows the total intensity map. The lower panel
    shows the polarization intensity map with contours of total intensities
    overlaid (the strong point-like sources listed in Table~2 have been
    subtracted). Contours start at 10\ mK~T$_{\rm B}$, and increase by a
    factor of $\sqrt{2}$. The bars show the orientation of the magnetic field
    (E+90$\degr$) for the case of negligible Faraday rotation.  Their length
    is proportional to the polarized intensity.}
\label{6cm}
\end{figure}

The Urumqi $\lambda$6\ cm total intensity map of S147 and polarization
intensity map with vectors overlaid in magnetic field direction are shown in
Fig.~\ref{6cm}. The total intensity map clearly reveals the spherical SNR
shell with a hollow region in its center. The r.m.s.-noise in the common area of all
coverages (see Table~\ref{obspara}) in total intensity
is 0.7\ mK~T$_{\rm B}$ and in polarized intensity is 0.3\ mK~T$_{\rm
B}$. These values are slightly larger than expected. It could be partially
caused by confusion. \citet{u+99} found a confusion limit of 15~mK~T$_{\rm
B}$ at 1400~MHz, which converts to 0.48~mK~T$_{\rm B}$ at $\lambda$6\ cm based on a
flux density spectral index of $\alpha = -0.9$ \citep{z+03}. Combining
this value with the theoretically expected value, we obtain an r.m.s.-noise of
0.5\ mK~T$_{\rm B}$. Another part may be explained by residual gain
fluctuations, ``scanning effects", atmospheric and ground radiation
variations, and low-level interference close to the noise level \citep{s06}.

\begin{figure}
\centering
\includegraphics[width=8cm]{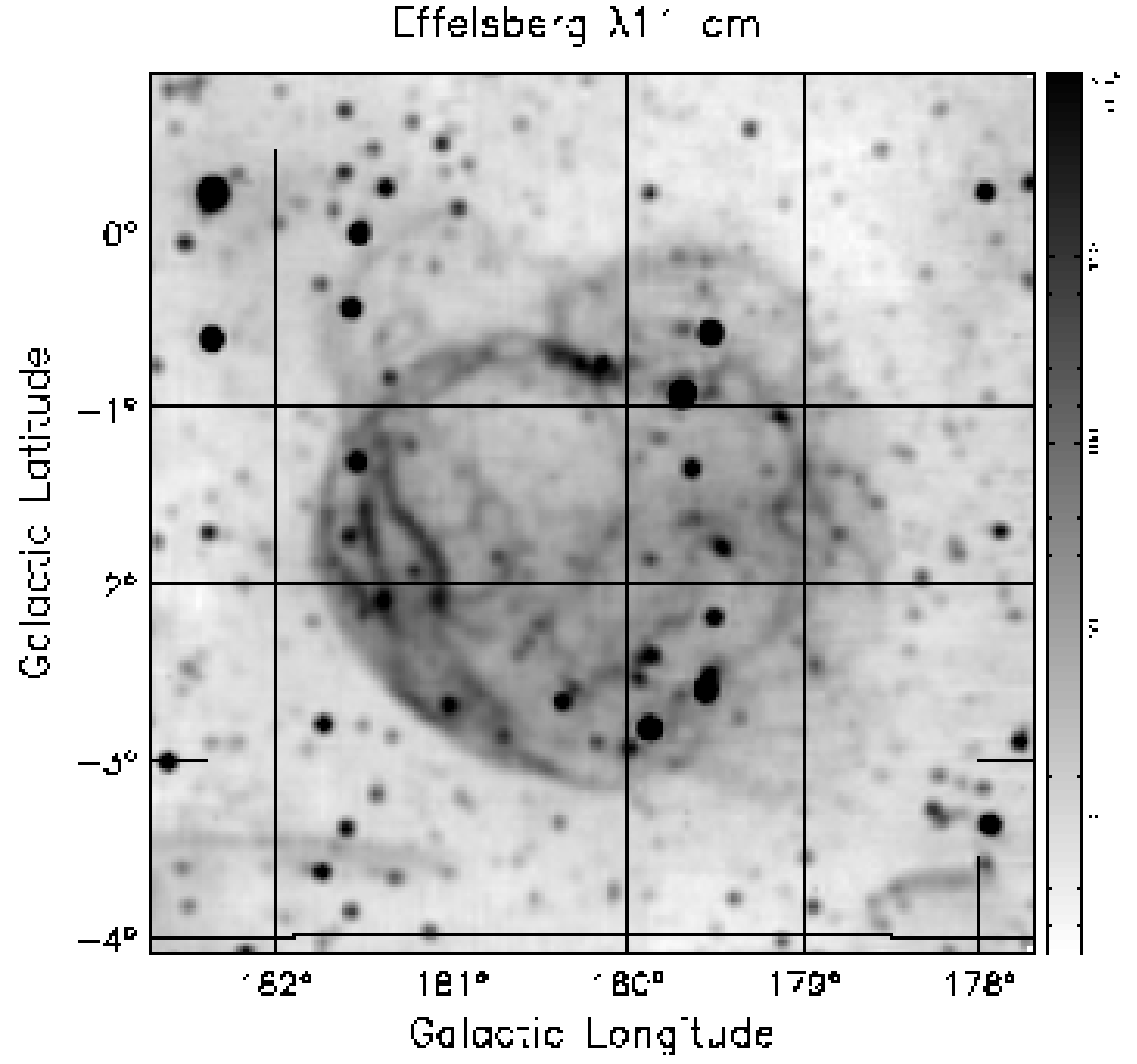}
\includegraphics[width=8cm]{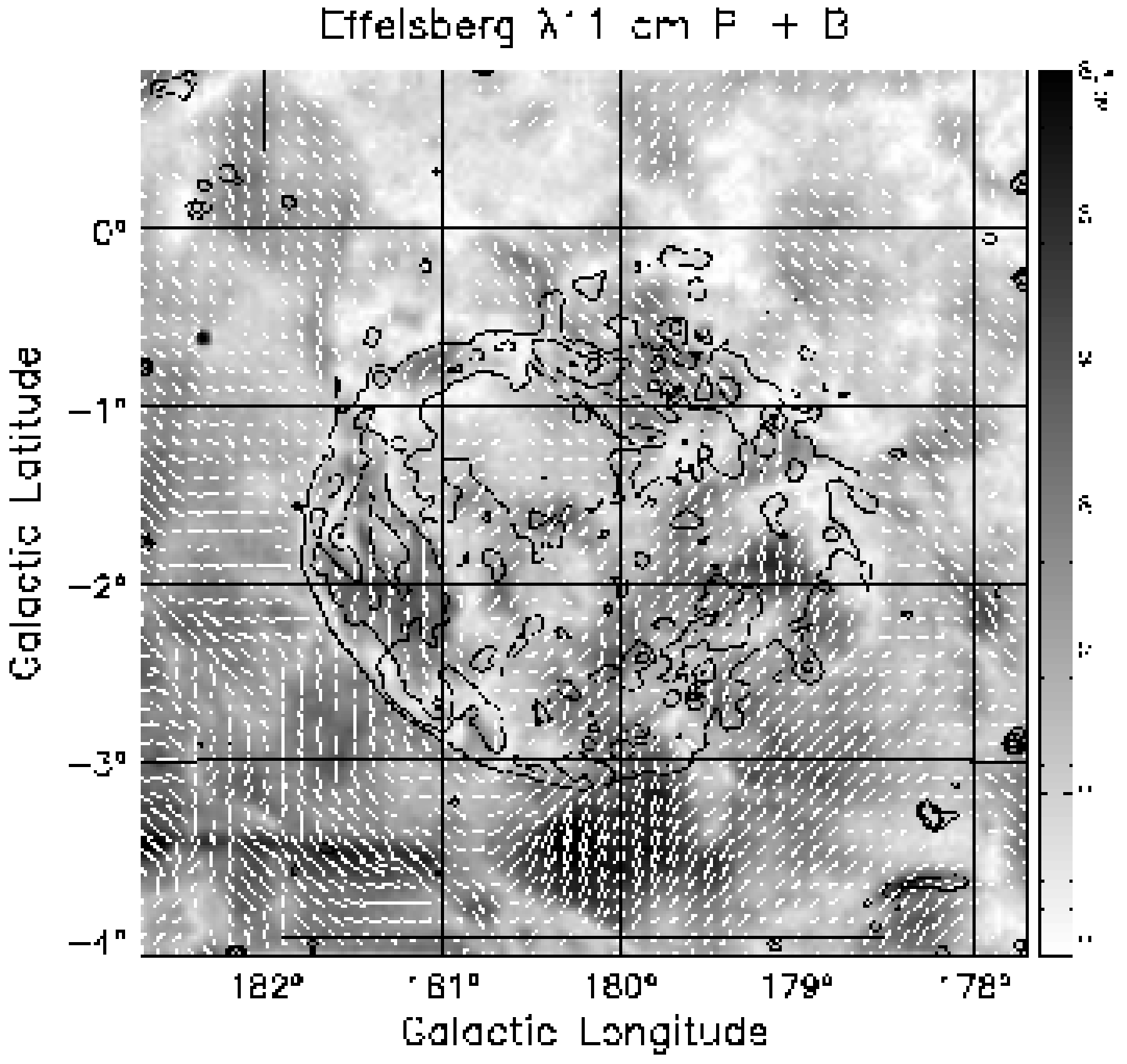}
    \caption{Same as Fig.~\ref{6cm} but showing the Effelsberg $\lambda$11\ cm maps at an angular
    resolution of 4\farcm4. Contours start from 50\ mK~T$_{\rm B}$ and increase by a
    factor of 2.}\label{11cm}
\end{figure}

\subsection{Observations at $\lambda$11\ cm}

We observed S147 at $\lambda$11\ cm with a new receiver installed in 2005 in the 
secondary focus of the Effelsberg 100-m radio telescope. 
The map size is $5\degr \times5\degr$ centered
at $l,b= 180{\fdg}2,-1{\fdg}6$. We made six coverages along Galactic
longitude and Galactic latitude direction with a scanning velocity of
$4\arcmin$/s. We processed the maps in the NOD2 format based standard
reduction package for observations with the Effelsberg 100-m telescope. Bad
data caused by strong interference were blanked. Scanning effects were suppressed 
by using the method of ``unsharp masking" developed by \cite{sr+79}. Finally, we 
``weaved" together all edited maps applying the method of \cite{eg+88}.

The $\lambda$11\ cm maps are shown in Fig.~\ref{11cm}. The r.m.s.-noise in
total intensity (upper panel) is measured to be 5.0\ mK~T$_{\rm B}$. In
polarized intensity, we found a value of 3.5\ mK~T$_{\rm B}$.  Because of the
higher angular resolution of 4\farcm4, the shell and the filamentary
structures can be more clearly identified than in the $\lambda$6\ cm map.

\begin{table*}
\centering
\caption{Strong point-like sources in the boundary of S147 detected by
Effelsberg at $\lambda$11\ cm and  Urumqi at $\lambda$6\ cm. The flux density
is given by the peak flux of two-dimensional gaussian fits.}
\label{sources}
\begin{tabular}{cccccrr}
\hline\hline
 Source & $\alpha_{1950}$ & $\delta_{1950}$ &  $\alpha_{2000}$ &  $\delta_{2000}$ & {$S_{11cm}$} &
 {$S_{6cm}$}  \\
     & (h~~m~~s)~~   & ($ ^\circ~~~\arcmin~~~\arcsec$)  & (h~~m~~s)~~ & ($ ^\circ~~~\arcmin~~~\arcsec$) &{(mJy)} &{(mJy)} \\
\hline
      0531+275  &05 31 18.3 &27 30 24 & 05 34 26.6 & 27 32 23 & 533$\pm$37 &287$\pm$44 \\
      0531+279  &05 31 21.7 &27 53 41 & 05 34 30.6 & 27 55 39 &383$\pm$33 &265$\pm$24 \\
      0533+282  &05 32 48.2 &28 09 24 & 05 35 57.5 & 28 11 16 &135$\pm$11 & 51$\pm$14 \\
      0533+272  &05 33 05.6 &27 10 43 & 05 36 13.5 & 27 12 33 &115$\pm$13 & 84$\pm$13 \\
      0534+284  &05 34 13.5 &28 24 41 & 05 37 23.2 & 28 26 26 &107$\pm$13 &130$\pm$20 \\
      0535+266  &05 34 33.1 &26 37 25 & 05 37 40.2 & 26 39 09 & 86$\pm$12 &     \\
      0536+285  &05 36 20.5 &28 30 19 & 05 39 30.4 & 28 31 56 &133$\pm$10 & 71$\pm$18 \\
      0538+287  &05 38 06.3 &28 41 14 & 05 41 16.5 & 28 42 43 &762$\pm$46 &527$\pm$47 \\
      0539+290  &05 39 02.7 &29 00 20 & 05 42 13.4 & 29 01 45 &665$\pm$42 &435$\pm$33 \\
      0541+269  &05 41 03.4 &26 55 07 & 05 44 11.0 & 26 56 23 &146$\pm$16 &      \\
      0544+273  &05 44 27.5 &27 20 58 & 05 47 35.7 & 27 21 59 &262$\pm$17 &251$\pm$22 \\
\hline
\end{tabular}\\
\end{table*}

\begin{figure*}
\centering
\includegraphics[width=7cm]{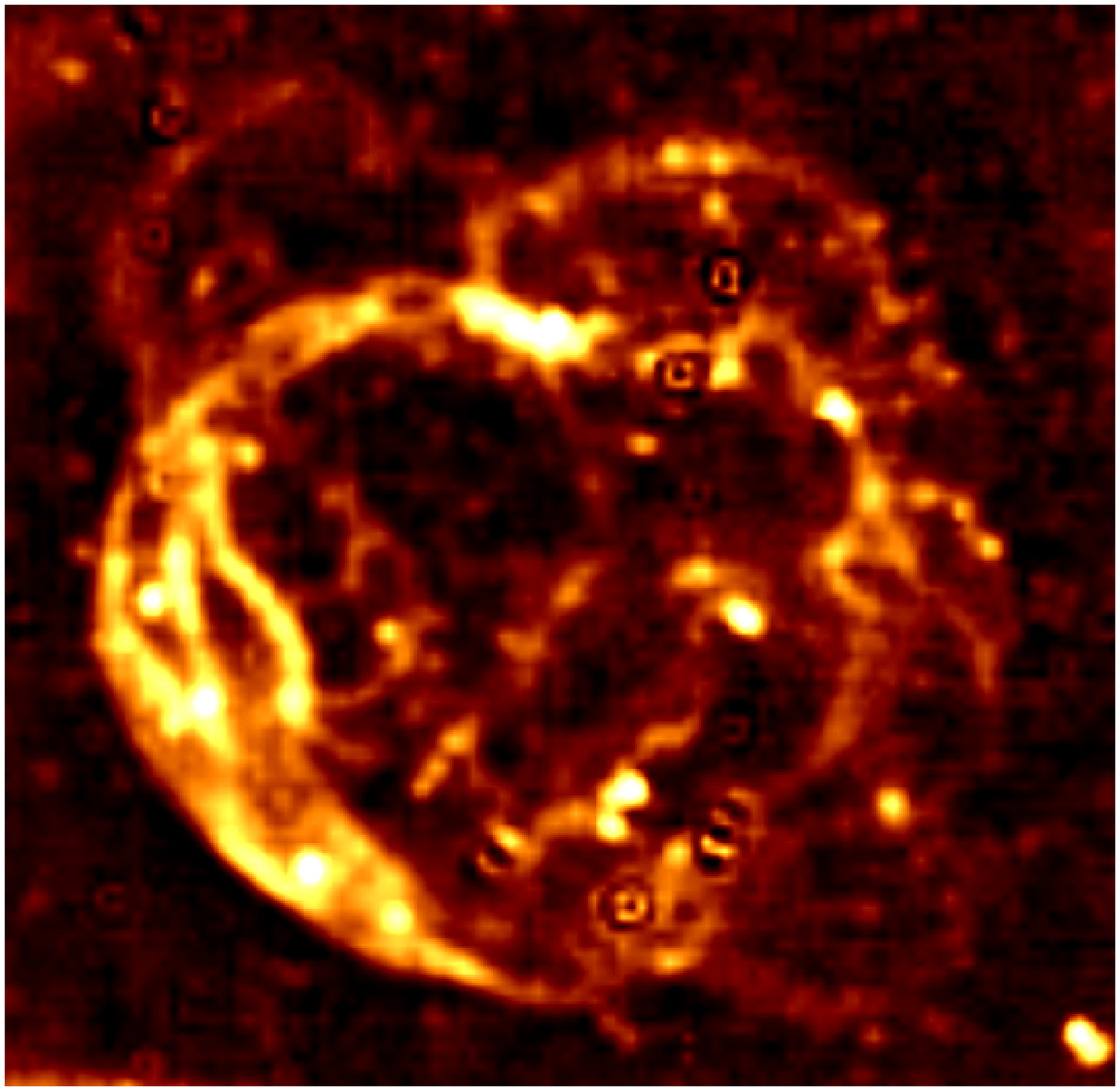}
\includegraphics[width=7cm]{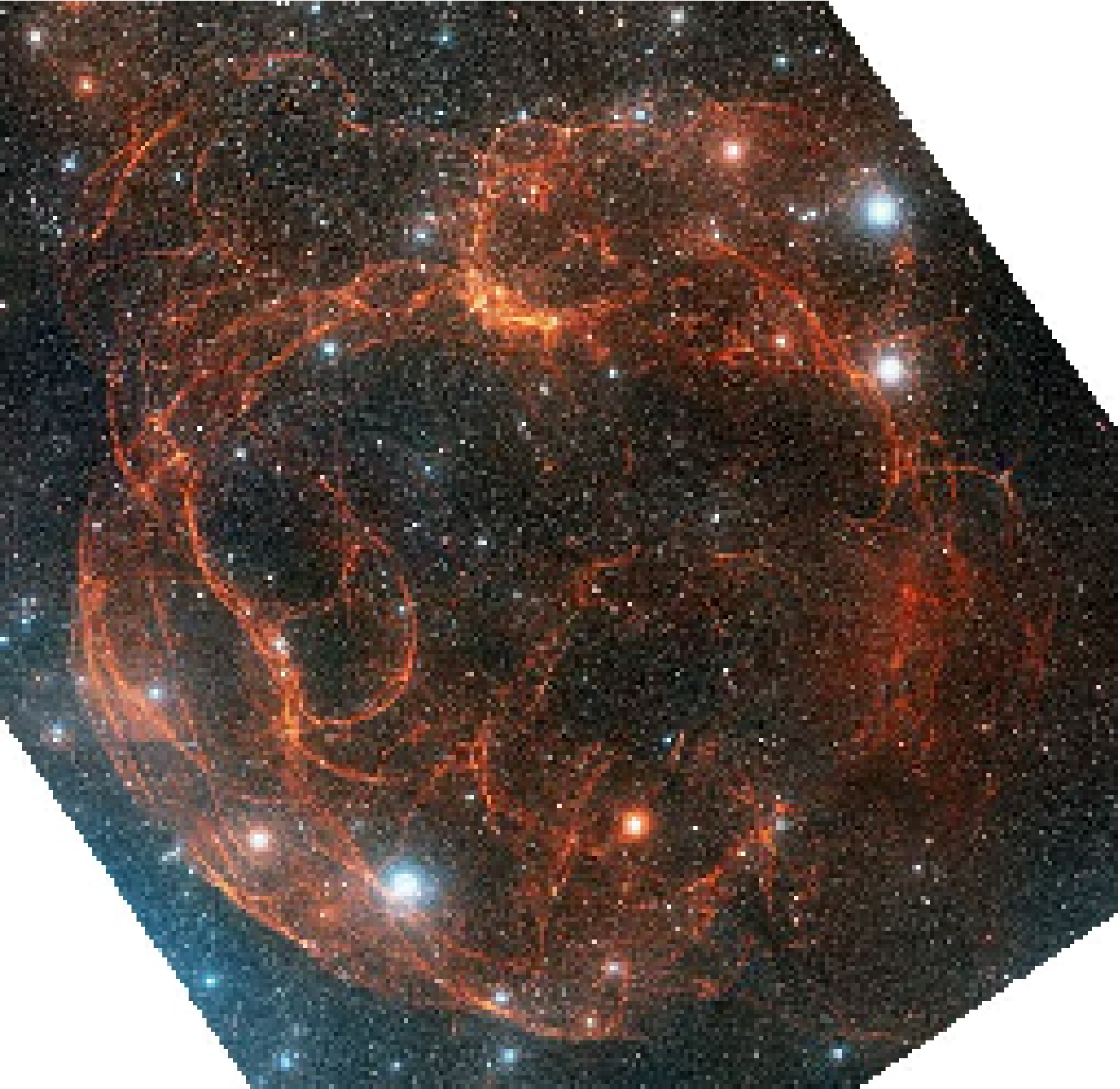}
   \caption{Left panel: The filamentary structure of S147 at $\lambda$11\ cm
    with most of strong point-like sources and diffuse large-scale ($>12'$)
    emission component subtracted (see text for details). Right panel: the optical image of S147 (Credit:
    Digitized Sky Survey, ESA/ESO/NASA,
    http://antwrp.gsfc.nasa.gov/apod/ap051129.html).}
    \label{filament}
\end{figure*}

\section{Comparison of radio and optical observations}

We used $\lambda$11\ cm total intensity map at the original angular resolution
after subtraction of the point-like sources listed in Table~\ref{sources}
to separate small-scale and large-scale structures applying
the ``background filtering" (BGF) procedure, invented by \citet{sr+79},
where a filtering beam of $12\arcmin$ was used.
A comparison of the small-scale radio structure with optical filaments of
S147 is displayed in Fig.~\ref{filament}, which demonstrates an excellent
agreement, even faint optical structures of the blow-outs toward north-east
and south-west are visible in both bands.

\begin{figure}
\centering
\includegraphics[width=8cm]{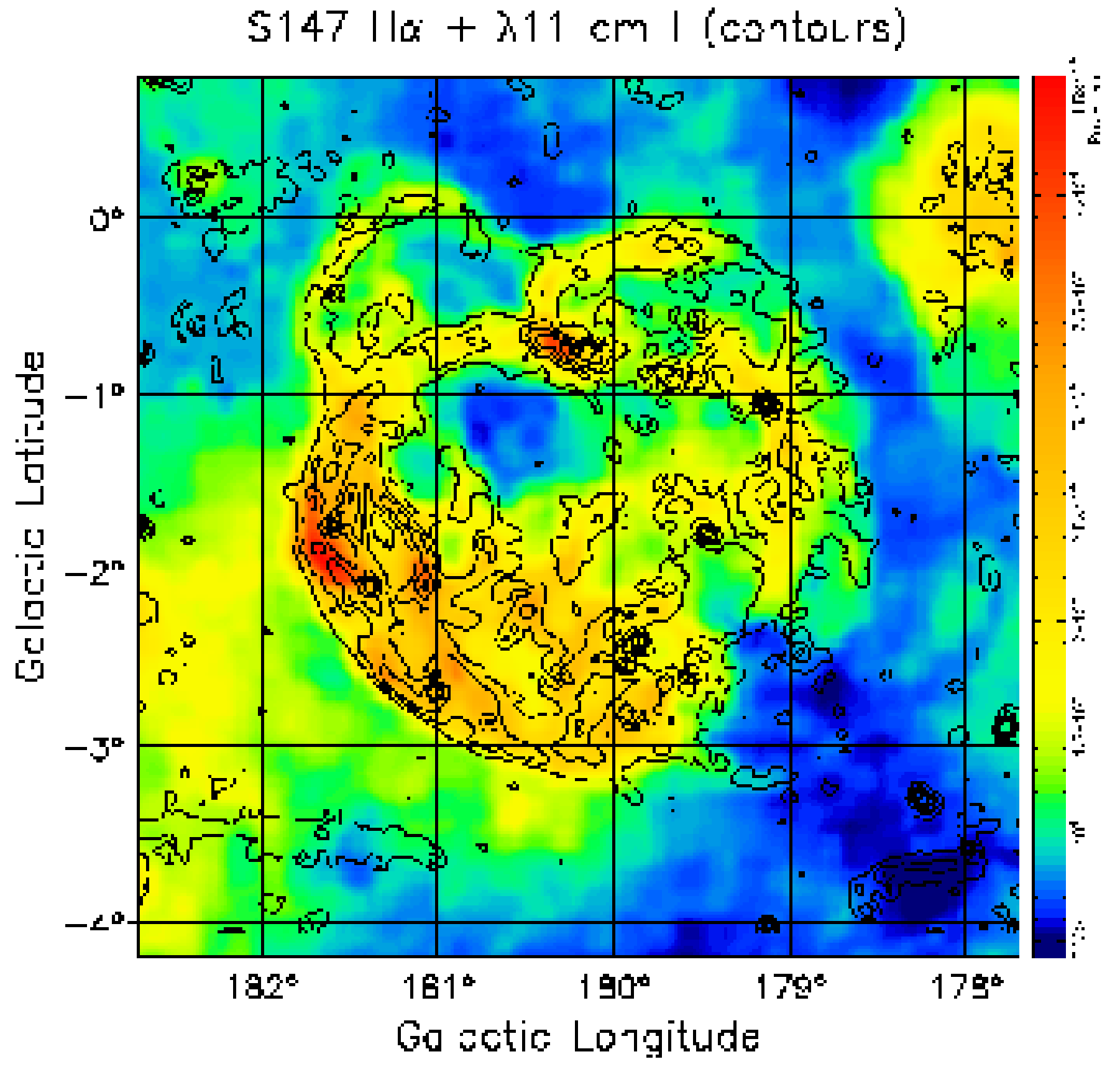}
    \caption{The filamentary structures of S147 observed at $\lambda$11\ cm
    is overlaid with H$_{\alpha}$ map \citep{f+03}. The angular resolution
    of the H$_{\alpha}$ map is 6$\arcmin$. The contours at $\lambda$11\ cm
    start at 20\ mK~T$_{\rm B}$ and run in steps of 35\ mK~T$_{\rm B}$.}
    \label{Halpha}
\end{figure}

We also overlay the radio filamentary structures on the H$_{\alpha}$ map
compiled by \citet{f+03} in Fig.~\ref{Halpha}, and find a rather good
positional coincidence when inspecting individual filaments. However, we find large
variations in the relative intensities, which have been already
mentioned by \citet{sf+80}. These variations may be explained by either
stronger magnetic fields in the radio bright filaments and/or lower
temperatures/densities in the optically-bright filaments. The measured
temperature and density variations \citep{f+85, d+77} support this
conclusion. \citet{d+77} obtained an average electron density of about 250\
cm$^{-3}$ assuming an electron temperature of $10^{4}$~K and showed that the
compression by an adiabatic shock is insufficient to obtain these
densities. Obviously, at least parts of S147 are already in the cooling
phase of SNR evolution.

\begin{figure}
\centering
\includegraphics[width=6cm,angle=-90]{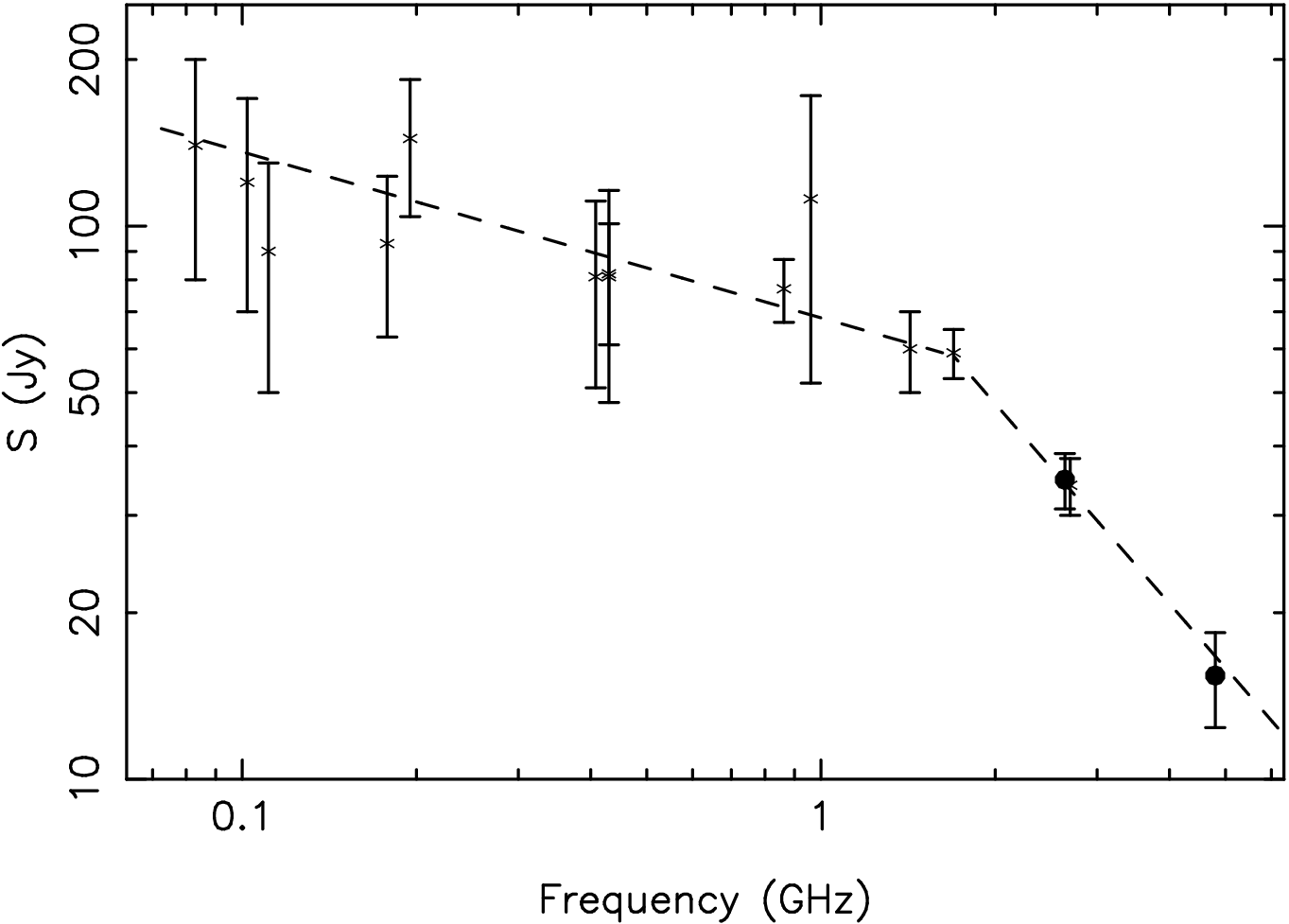}
    \caption{Spectrum of the integrated radio flux densities of S147. The
    low-frequency spectrum has a spectral index of $\alpha \sim -0.30\pm
    0.15$. Above the break the spectral index is  $\alpha \sim -1.20\pm
    0.30$. The S147 integrated flux densities are taken from \citet{fr+86}; \citet{rzf+03}; and \citet{kpu+94}.
    We omitted the $\lambda$6\ cm flux density value obtained by \citet{sf+80}, because this
measurement did not cover the entire source.
    \citet{rzf+03} have corrected the values for the contribution of
    point-like sources as listed by \citet{f+82}, 
    assuming a mean source spectral index of $\alpha=-0.75$ at frequencies lower than 408\ MHz.}
  \label{spectral}
\end{figure}

\section{Spectral properties of S147}

\subsection{The integrated radio spectrum of S147}

The new observations with the Urumqi 25-m radio telescope provide the first
complete map of S147 at $\lambda$6\ cm. In order to estimate the integrated
flux density of S147, we subtracted eleven strong point-like sources (see
Table~\ref{sources}) toward S147 from the $\lambda$6\ cm and $\lambda$11\
cm maps. The flux densities of these sources have previously been listed by \citet{f+82},
and agree within 0.1~Jy to those in Table~\ref{sources}. We calculated the integrated 
flux density with the method of
ring integration by summing up the emission in concentric rings centered on
$l,b=180{\fdg}2,-1{\fdg}6$.
The integration stopped at a radius of
2{\fdg}0. Beyond this radius, we found an average base-level of $\rm 8.7\ mK\
T_{\mathrm B}$ at $\lambda$11\ cm and of $\rm 1.9\ mk\ T_{\mathrm
B}$ at $\lambda$6\ cm. We have subtracted these base-levels from the maps
before calculating the flux density of S147. Taking these local base-levels
into account, polygon integrations  at the two wavelengths just outside 
the periphery of S147 result in the same flux density values as obtained from the method
of ring integration. From variations just outside S147 we found the uncertainty 
in the base-levels to be $\rm 2\ mk\
T_{\mathrm B}$ at $\lambda$11\ cm and $\rm 1\ mk\ T_{\mathrm B}$ at
$\lambda$6\ cm. Considering 5\% of other uncertainties including
calibration, the flux density values are $\rm 34.8\pm 4.0\ Jy$ at $\lambda$11\
cm and $\rm 15.4\pm 3.0\ Jy$ at $\lambda$6\ cm.  Figure~\ref{spectral} displays
the integrated spectrum of S147 including available low-frequency flux
density data.

\citet{rzf+03} placed the turnover frequency between 1.5\ GHz and 2\ GHz
using the extrapolated flux density of the incomplete 5\ GHz map observed by
\citet{sf+80}. The extrapolated flux density value as well as the spectral
turnover at $\sim1.5$\ GHz are clearly confirmed by new measurements at
$\lambda$6\ cm and $\lambda$11\ cm. The spectrum is flat at lower
frequencies and steepens toward frequencies higher than 1~GHz to 2~GHz.
The low-frequency spectrum has a spectral index of about $\alpha \sim
-0.30\pm 0.15$. The spectral index is about $\alpha \sim -1.20\pm 0.30$ above
the break frequency.
The relatively large error at high
frequencies is mainly caused by the still large uncertainty of the flux
density at 4800\ MHz ($\lambda$6\ cm), which, in principle, is very difficult to reduce for
faint large diameter objects like S147 located in the Galactic plane.

\subsection{The spectral index map of S147}

The Urumqi $\lambda$6\ cm and the Effelsberg $\lambda$11\ cm maps with point-like sources and
constant base-levels removed  were
both convolved to a common angular resolution of $10\arcmin$. We estimated the relative
position accuracy of the two maps by using the positions of
point-like sources from Table~2. We found the mean positional difference to
be $6\arcsec$ with a maximum of $28\arcsec$.

\begin{figure}[!htbp]
\centering
\includegraphics[width=8cm]{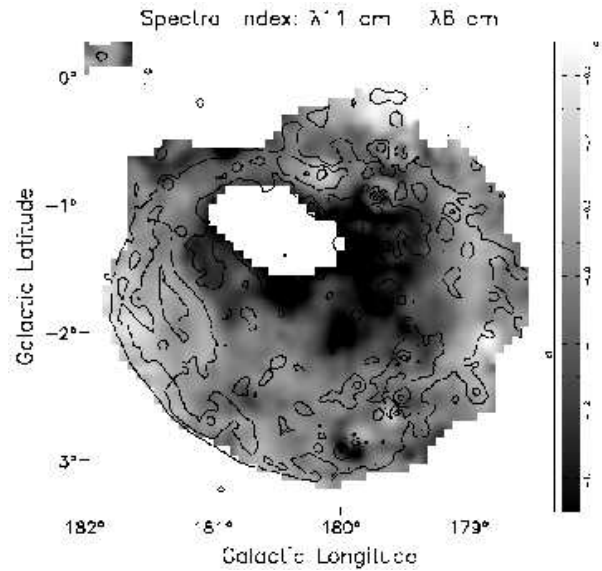}
\caption{Spectral index map calculated between $\lambda$11\ cm and $\lambda$6\ cm at
10$\arcmin$ angular resolution.  The overlaid contours represent total
intensities at $\lambda$11\ cm (with point-like sources listed in Table~2
subtracted), starting from 50\ mK~T$_{\rm B}$ and increasing by a factor of
2.}
\label{11-6}
\end{figure}

We display the spectral index map between $\lambda$11\ cm and
$\lambda$6\ cm in Fig.~\ref{11-6}. We calculated the spectral index of each pixel from 
the brightness temperatures at the two frequencies. In order to achieve accurate
spectral indices and to exclude the influence of systematic effects we set a
lower flux density limit of 40\ mK~T$_{\rm B}$ and 6\ mK~T$_{\rm B}$ for the
$\lambda$11\ cm and $\lambda$6\ cm map, respectively. The possible variations of the base-levels
at $\lambda$11\ cm and $\lambda$6\ cm cause an uncertainty of the spectral indices of $\Delta \alpha \sim$ 0.3. 
The uncertainty is largest where the total intensity is small. The general result of a much flatter radio
spectrum  of the filaments compared with the large-scale component is not altered. However, systematic errors 
make it difficult to derive positional spectral index variations of the large
scale component. The spectrum for
regions with bright radio filaments, particularly in the eastern and
northern part of S147, is much flatter than for regions with less pronounced
filaments and for the large-scale diffuse emission component near the center.
 As seen in the total intensity maps, faint, extended Galactic
emission exists outside of S147. Such emission features may also
exist in the projected region toward S147, leading to variations of the spectral
index.

\subsection{T-T plot analysis}

\begin{figure*}
\centering
\includegraphics[width=8cm,angle=-90]{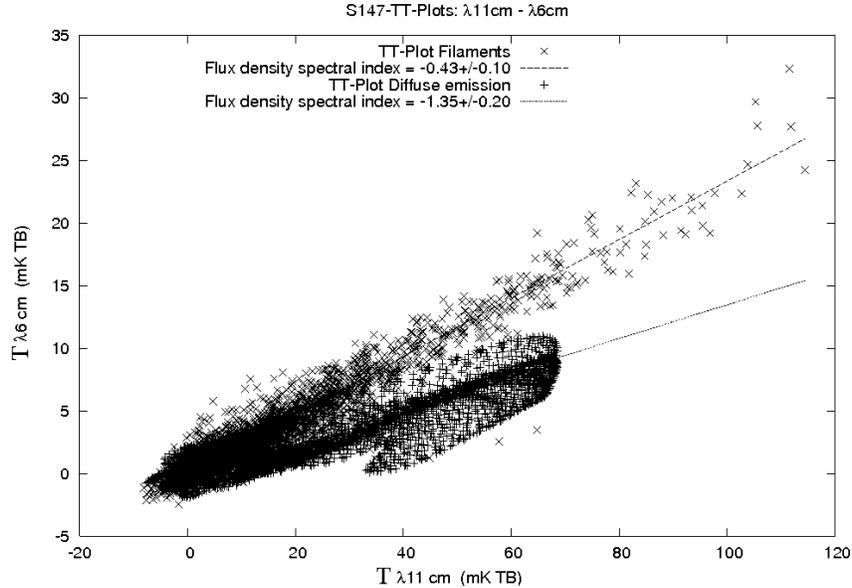}
\caption{The T-T plot for the filamentary and diffuse emission components of S147
 between $\lambda$11\ cm and $\lambda$6\ cm. Filamentary emission component
 in general has a spectral index of $\alpha= -0.43\pm 0.10$. The diffuse
 emission component has a spectral index of $\alpha= -1.35\pm 0.20$.}
 \label{fila-diff}
\end{figure*}

The temperature$-$temperature (T-T) plot is a method \citep{tp+67} to
investigate the spectral behavior, independent from a constant base-level
across the source.
We used the maps at an angular resolution of 10$\arcmin$ at $\lambda$11\ cm
 and $\lambda$6\ cm to check the spectral index of the filamentary and diffuse
emission components. We first decomposed both maps into small-scale filamentary
structures and large-scale diffuse emission component with the BGF procedure
described by \citet{sr+79} with a filtering beam of $18\arcmin$ and
cut-values of 100~mK~T$_{\rm B}$ at $\lambda$11\ cm and 25~mK~T$_{\rm B}$ at
$\lambda$6\ cm. The resultant TT-plots for the filaments and for the diffuse
emission component are shown in Fig.~\ref{fila-diff}. They have largely different
spectral indices.  We found the steep spectrum with a spectral index of
$\alpha=-1.35\pm 0.20$ for the diffuse emission component, while we found 
a spectral index $\alpha=-0.43\pm 0.10$ for the filamentary
emission component. This index is slightly steeper than the integrated low-frequency
spectrum ($\alpha = -0.30\pm0.15$), but is comparable considering the quoted
uncertainties. The quoted errors include the statistcal error (number of data
points reduced to one per HPBW) as well as the error, which follows from fitting the
data twice, alternatively taking the data of one of the two wavelengths as independent variable.
The result in Fig.~\ref{fila-diff} may be affected by possible variations of the base-levels across
S147. This was inspected using TT-plots of four smaller regions separated by l=$180{\fdg}0$ and
b=-$1{\fdg}9$. For the filamentary component, the variation of the spectral index is within 
the quoted error of $\Delta \alpha=\pm$0.10.
Therefore, the filamentary emission component probably has a fairly
straight radio spectrum over the accessible frequency range. 
For the diffuse emission component the spectral index varies between $\alpha$=-1.10 and
$\alpha$=-1.56. There is no doubt that the radio spectrum of the diffuse emission component is much
steeper than that of the filamentary component. This is in agreement with the result from the
spectral index map. It is entirely the
diffuse emission component that produces the spectral turnover around 1.5~GHz. The
integrated flux densities of the separated diffuse and filamentary emission components
are $\sim 22.8$~Jy and $\sim 12$~Jy at $\lambda$11\ cm, and $\sim 8.4$~Jy and $\sim
7$~Jy at $\lambda$6\ cm, respectively, which also shows the difference of spectral
indices of the two components.

It is clear that above 4800\ MHz the flux density of the radio filaments
dominates the integrated flux density of S147, causing a spectral flattening
toward higher frequencies. To check this, we inspected the WMAP data in all
five bands between 22.8\ GHz and 93.5\ GHz \citep{h+07} for a signature of
S147. We are able to
clearly trace S147 up to the highest frequency especially its region with
strong radio filaments near $l=181\degr$.

\begin{figure}
\centering
\includegraphics[width=8cm]{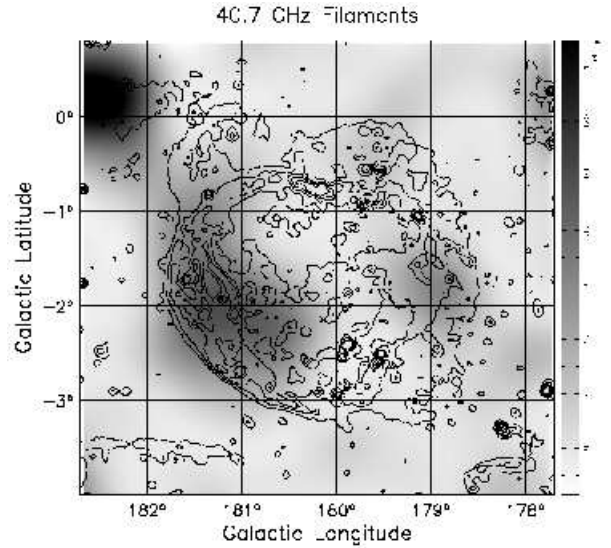}
\caption{The filamentary emission component of S147 at 40.7\ GHz (grayscale)
observed by WMAP at an angular resolution of 49$\arcmin$. $\lambda$11\ cm 
contours of Effelsberg measurements at an angular resolution of $4\farcm4$
start at 20\ mK~T$_{\rm B}$ and run in steps of 35\ mK~T$_{\rm B}$.}
\label{41fila}
\end{figure}

\begin{figure}
\centering
\includegraphics[width=4.5cm,angle=-90.0]{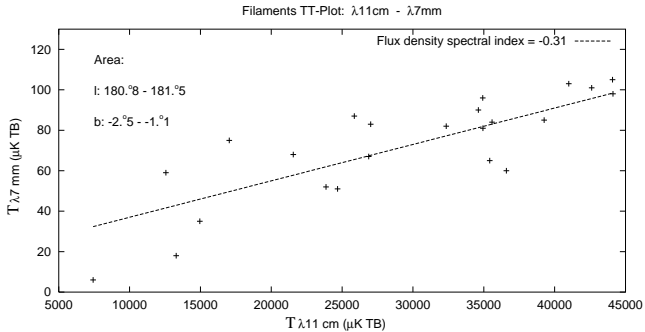}
\caption{The T-T Plot of filamentary emission component at $\lambda$7\ mm (40.7\ GHz) and $\lambda$11\ cm.
The spectral index is $\alpha=-0.31 \pm 0.20$.}
\label{fila_41_2639}
\end{figure}

We convolved all WMAP data and the $\lambda$11\ cm map to the
beam size of the 22.8~GHz map ($49\arcmin$).
We separated the filamentary emission component of S147 in all maps 
by applying the BGF procedure described by \citet{sr+79}, using a filtering
beam size of $50\arcmin$. As an example, we show the WMAP data at 40.7~GHz
in Fig.~\ref{41fila} at $49\arcmin$ angular resolution in comparison with
the filamentary emission component of the Effelsberg $\lambda$11\ cm  map at the original
angular resolution. To check the spectral index, we performed a T-T Plot
(Fig.~\ref{fila_41_2639}) for the filamentary emission component at 40.7~GHz and $\lambda$11\ cm
 of the area $l=180{\fdg}8$ to $l=181{\fdg}9$, $b=-0{\fdg}87$ to
$b=-2{\fdg}7$. We found the spectral index to be $\alpha = -0.31\pm 0.20$,
which is consistent with that between $\lambda$11\ cm and $\lambda$6\ cm. A similar
spectral index for the filamentary emission component is obtained between $\lambda$11\ cm
and 22.8\ GHz and 33.0\ GHz, respectively.

From the various methods we may derive a spectral index for the filamentary emission component
of  $\alpha = -0.35\pm 0.15$.
Using the flux density of 7~Jy for the filamentary
structures at $\lambda$6\ cm, we calculate the flux density of
filamentary emission component to 3.5~Jy at 40.7\ GHz, 4.3~Jy at 22.8\ GHz, and 3.8~Jy
at 33.0\ GHz. The flux density of the diffuse emission component of S147 at 40.7\ GHz or
other WMAP frequencies can not be determined since a separation of this
component from the underlying stronger large-scale Galactic emission is not
possible.

At 60.8\ GHz and 93.5\ GHz the emission in the quoted area increases
significantly, which is most likely due to thermal dust emission associated with the SNR.

\begin{figure}
\centering
\includegraphics[width=8cm,angle=-90]{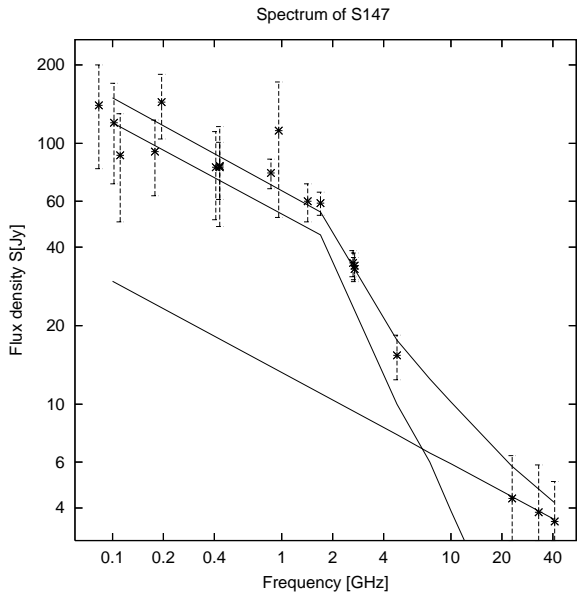}
\caption{Integrated total intensity spectrum of S147, including WMAP data for
the filaments. The weaker contribution from the diffuse emission component can not be
obtained from the WMAP data. The model spectrum of the filamentary
component, the diffuse component with a spectral break at 1.7~GHz, and the
sum of both are shown in addition (see text).}\label{spectral2}
\end{figure}

We show all flux densities up to 40.7~GHz in Fig.~\ref{spectral2}, where the
WMAP values miss the small contribution from the diffuse emission
component. In addition, we have plotted the supposed straight spectrum of the
filamentary component with a spectral index of $\alpha = -0.35$. We have
also plotted the contribution of the diffuse large-scale emission component and the
sum of both.  The turnover of the integrated total intensity
spectrum has to be attributed to the turnover of the large-scale emission
component. We assume that both emission components have the same spectral
index below the break frequency. In summary, the integrated total intensity
spectrum is flat at low frequencies, steep between 1~GHz and about 10~GHz,
and flattens again toward higher frequencies. No other SNR is known showing
a similar radio spectrum. Hence, the radio spectrum of S147 is unique.

\subsection{The analysis of the spectral turnover}

A spectral turnover is very rare for shell-type SNRs. Some early claims
have turned out to result from limited accuracy of high-frequency
measurements. S147 is the best known case for a spectral break so far.

Several possible mechanisms cause a spectral turnover: (1) The presence of
two populations with different energy spectra; (2)
shock acceleration of particles; (3) the compression of the Galactic
magnetic field; and (4) synchrotron losses (aging) of electrons.

(1), (2) According to \citet{lr+98} the first two mechanisms may not be relevant
for a synchrotron spectrum as given in Fig.~\ref{spectral}.

(3) The Galactic relativistic electron spectrum has a slow turnover due to
various loss and propagation effects over time. \citet{w+74} has compiled
the Galactic radio spectrum, which varies from $\alpha \sim$-0.5\ to $\alpha
\sim$-0.8\ between $\sim$100\ MHz and $\sim$600\ MHz.  The corresponding electron
energy index varies from $\sim$2.0\ to $\sim$2.6\ over the energy range
2.6~GeV to 6.5~GeV in case the magnetic field strength is 3~$\mu$G. Direct
observations of the electron spectrum \citep{t+84} also reveal a change of
the Galactic electron spectral index from 2.0 at low energy ($< 2$~GeV) to
3.5 at higher energies. Due to the shock wave of the SNR, the compressed
local magnetic field in a SNR shifts the spectral turnover toward
a higher frequency, e.g., that at about 1.5~GHz of S147.
\citet{d+74} has studied this case extensively. A crucial parameter
is the radio flux density, which is related to the observed turnover frequency
and the distance. Following the argument by DeNoyer, it is possible
to explain the observed radio spectrum of S147, in particular, for the smaller distance
of $\le$1~kpc.

(4) If synchrotron losses of electrons occur uniformly over the whole lifetime
of the SNR, the turnover would happen at $\rm \nu_{c}(MHz)=3.4\times
10^{9}~B(10\mu G)\, \mbox{}^{-3}~t(10^{4}yr)\, \mbox{}^{-2}$. For S147 the
turnover frequency $\nu_{c}$ is observed to be at 1.5~GHz and its age is
about $10^{5}$yr, so the magnetic fields should be about 0.3~mG. Such
a high magnetic field is not known to exist in shell type SNRs.
\citet{d+74} discusses the situation that the acceleration and synchrotron
losses occur during the early phase of SNRs. \citet{g+73} has shown that in
the early phase Rayleigh-Taylor instabilities create a convective zone,
which can sustain magnetic fields B of 1000\ $\mu$G within young remnants
for approximately 100~years, corresponding to a radius $R_c$ of the remnant
of $\sim 1$\ pc in the low-density region of the Galactic anti-center.
Due to the subsequent expansion, the magnetic field decreases with
(R/R$_c$)$^{-2}$, while the energy decreases with (R/R$_c$)$^{-1}$. The
turnover frequency decreases with (R/R$_c$)$^{-4}$. For S147, R/R$_c$ is 21
for a distance of 800\ pc corresponding to a turnover frequency of 12\ MHz
and 154\ MHz, respectively.  However, as \citet{d+74} pointed out, at this stage
of evolution the magnetic field of the SNR is dominated by the shock
compressed interstellar magnetic field, which may be about 10\ $\mu$G in
the large-scale emission area. In that case, the turnover frequency decreases
with (R/R$_c$)$^{-2}~\times$~(10~$\mu$G/1000~$\mu$G) to about 188\ MHz and
680\ MHz, respectively.  Given the uncertainty of the magnetic field strength
during the synchrotron loss phase, the interstellar magnetic field 
in the area of S147, as well as
the amount of shock compression (with or without cooling effects), this model cannot be
ruled out as a possible explanation of the observed spectral bend.

We conclude, in principle, both the compressed Galactic magnetic field
scenario and early high synchrotron losses with subsequent adiabatic
expansion could explain the observed turnover. The observed difference of
$\Delta \alpha \sim 1$ between the high-frequency and the low-frequency
spectral indices is larger than the value given by \citet{t+84}, but smaller
than that from synchrotron losses with no further injection of relativistic
electrons after the first 100 years \citep[see Fig.~2.12 in][]{p+77}.

\section{The polarization towards S147}

$\lambda$6\ cm and $\lambda$11\ cm polarization maps are shown 
in the lower panels of Figs.~\ref{6cm} and ~\ref{11cm}. The polarization features 
seen in the two polarization maps
are complex, but very similar at both wavelengths. They have no counterpart
in the total intensity map except for some sections of the S147 shell.
Obviously, the interstellar diffuse emission component is responsible for most of the
observed polarization features. The similarity of polarized structures at
both wavelengths suggests that Faraday rotation along the line of sight is
very small.

\begin{figure}
\centering
\includegraphics[width=8cm]{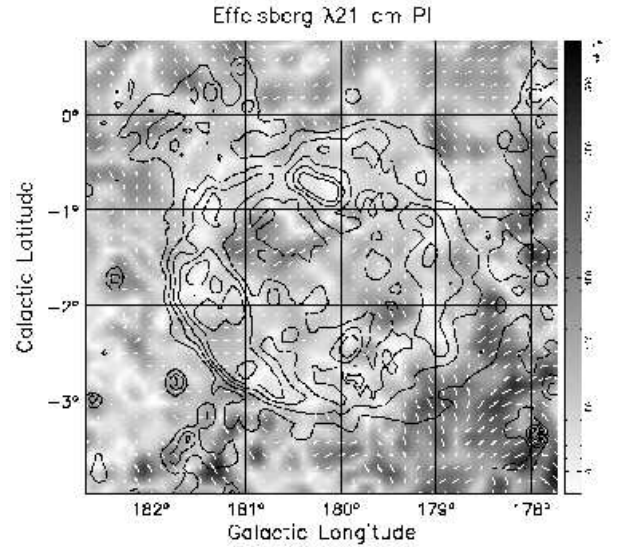}
\caption{$\lambda$21\ cm polarization intensity map of S147 from the
Effelsberg 100-m telescope at an angular resolution of $9\farcm4$ (grey scale). The
r.m.s.-noise is $\rm 8\ mK\ T_{\mathrm B}$.  Contours show total intensities
(the strong point-like sources were subtracted) starting from 150\
mK~T$_{\rm B}$ and increasing {\bf in steps of 100\ mK~T$_{\rm B}$}. The bars show the
orientations of (E+90$\degr$). The length of the bars is proportional to the
polarized intensity.}
\label{21cm}
\end{figure}

We extracted the polarization intensity map of the same region
(Fig.~\ref{21cm}) from the $\lambda$21\ cm ``Effelsberg Medium Galactic
Latitude Survey" (EMLS) \citep{rf+04}. This map also includes the large-scale
polarized emission component at an absolute zero level from the polarization survey
made with the DRAO 26-m telescope \citep{wlrw+06}. The polarized
emission  along the bright shell of S147  is very low.
The shell of S147 might contribute
polarized emission, which adds to the Galactic emission  forming minimum polarization. 
As an alternative S147 might depolarize the emission from larger distances.

Large-scale emission component is missing in the polarization data at $\lambda$6\ cm
and $\lambda$11\ cm, which is known to severely limit any
interpretation of polarization structures caused by Faraday rotation effects
in the interstellar medium \citep{r+07}. Following \citet{s+07} we used the
WMAP polarization data at 22.8\ GHz to recover the missing large-scale
structures at $\lambda$11\ cm and $\lambda$6\ cm.

For this purpose, we convolved the WMAP U and Q maps to 49$\arcmin$ and
scaled to $\lambda$11\ cm and $\lambda$6\ cm with a  temperature spectral
index of $\beta = -2.8$, equal to the temperature spectral index toward the
Galactic anti-center \citep{rr+88}. We obtained a spectral index
of the polarized emission toward S147 of $\beta = -3.0\pm0.4$ from a T-T plot
between the $\lambda$21\ cm map and the WMAP 22.8\ GHz map, which are both
on an absolute scale. Although the uncertainty is large, the spectrum
of the polarized emission is consistent with that for total intensity, implying
that the polarized emission at longer wavelengths is not much reduced by
depolarization and that a low Faraday rotation occurs in the direction of
the Galactic anti-center. This justifies the usage of the better known
temperature spectral index $\beta = -2.8$.

The difference of the scaled WMAP data and the $\lambda$11\ cm and
$\lambda$6\ cm maps, which were convolved to the same beam, are then added
to the original U and Q maps at $\lambda$6\ cm and $\lambda$11\ cm
wavelength.  This procedure needs, however, to be modified in case of
significant Faraday rotation along the line of sight in particular for the
$\lambda$11\ cm map. We have also used the $\lambda$21\ cm (Fig.~\ref{21cm}) 
convolved to a 49$\arcmin$ beam and the WMAP 22.8\ GHz map to obtain the
average Faraday rotation over the entire area. We obtain a value of $\rm
\sim +10~rad~m^{-2}$. A low Faraday rotation is also supported by the pulsar
PSR J0538+2817, associated with S147, which has a Faraday rotation of $-7\pm
12$\ rad m$^{-2}$ \citep{m+03}. We conclude that the correction of the
polarization angle at $\lambda$11\ cm is smaller than a few degrees.  In the
following, we assume that no modification of the straightforward procedure is
necessary.

\begin{figure}
\centering
\includegraphics[width=8cm]{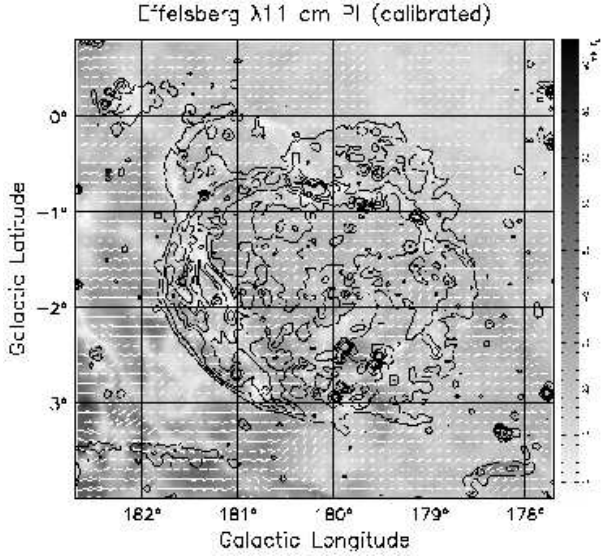}
\caption{Polarized $\lambda$11\ cm emission in the area of S147
after restoration for an absolute zero-level using WMAP 22.8\ GHz data  (grey scale).
The angular resolution is $4\farcm4$. Contours show total intensities
from the Effelsberg $\lambda$11\ cm map (with strong point-like sources
subtracted). The bars show the orientation of the magnetic field
(E+90$\degr$) in case of ignorable Faraday rotation. The length of
the bars is proportional to the polarized intensity.}
\label{11abs}
\end{figure}

\begin{figure}
\centering
\includegraphics[width=8cm]{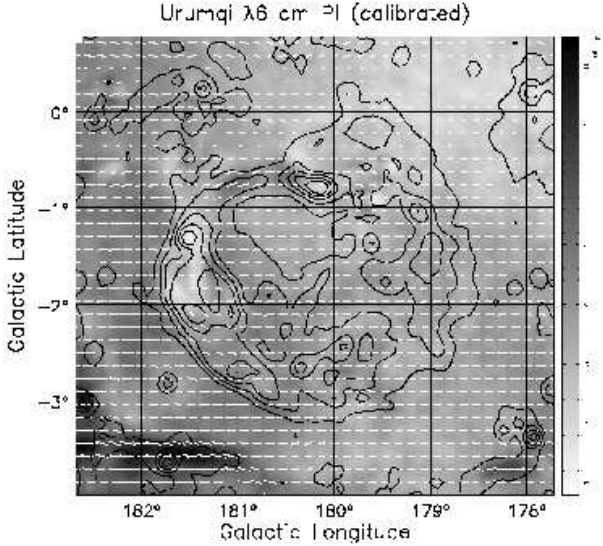}
\caption{Same at Fig.~\ref{11abs} but for  Urumqi $\lambda$6\ cm map.
The angular resolution is $9\farcm5$.}
\label{6abs}
\end{figure}

As expected for a magnetic field orientation along the Galactic plane a
large offset is found for Stokes Q (average $\sim$ 4\ $\rm mK~T_{B}$ at
$\lambda$6\ cm), while the offset for Stokes U is small (average
$\sim$ -0.2\ mK $\rm T_{B}$ at $\lambda$6\ cm). The polarized intensity maps
at $\lambda$11\ cm and $\lambda$6\ cm, after restoration for an absolute zero
level, are shown in Figs.~\ref{11abs} and \ref{6abs}. The strongest polarized
emission is located outside the boundary of S147 in both maps. The
distribution of polarized intensity in the $\lambda$6\ cm map does not allow
the identification of S147 in general. Only a few filaments,
e.g., $l,b =181{\fdg}6,-1{\fdg}0$ and $l,b =181{\fdg}2,-2{\fdg}0$, show
an association between total intensity filaments and  depressions in
polarized intensity. In the $\lambda$11\ cm map, several filaments are
clearly visible as depressions in polarized intensity. At both frequencies,
the appearance of S147 against the large-scale polarized emission component of the
Galactic background is similar to its appearance at $\lambda$21\ cm (see Fig.~\ref{21cm}),
but the ``depolarization" effect is less serious at higher frequencies.

The interpretation of the restored polarization data requires modeling of
the diffuse Galactic polarized emission component including Faraday
rotation effects \citep{r+07}.  However, the strong eastern filamentary structure is
a region where polarized emission from the SNR adds to the 
Galactic emission. It produces a depression in the polarized intensity map because
the magnetic field along the filaments is oriented almost
perpendicular to the general Galactic magnetic field direction. The strong polarized
filament structure is visible in the original polarization map at a relative
zero level (Fig.~\ref{6cm} and Fig.~\ref{11cm}).

\subsection{Rotation measures of the eastern filaments}

\begin{figure}[!htbp]
\centering
\includegraphics[width=6cm]{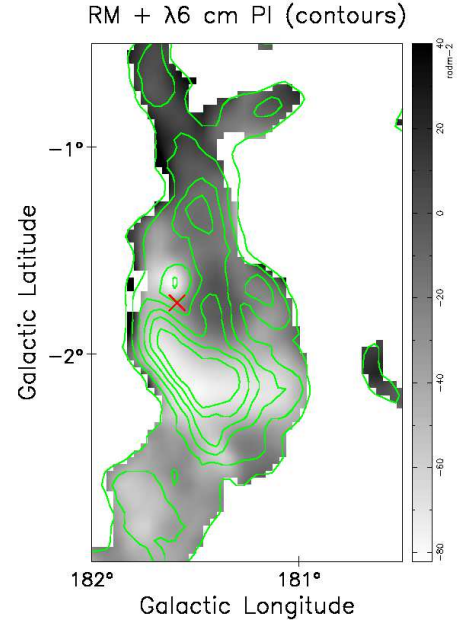}
\caption{The RM distribution for the eastern filamentary emission region of S147
derived from the original $\lambda$11\ cm and $\lambda$6\ cm polarization
maps. Contours show polarized intensities from the Urumqi $\lambda$6\ cm map
starting at 2.5~mK~T$_B$ and running in steps of 0.8~mK~T$_B$. 
The position of the extragalactic source 0539+266B with a RM of $-75\pm 4$\ rad m$^{-2}$ 
\citep{k+88} is indicated.}\label{rm}
\end{figure}

For the calculation of rotation measures (RMs) of S147 we have to remove the
diffuse Galactic polarized emission component. We noticed that the polarized emission from
the eastern filament is strong and outstanding against the background in the
original polarization map. We calculated the RM map from the polarization
angles of the original $\lambda$11\ cm and $\lambda$6\ cm polarization maps
after both maps have been convolved to a common angular resolution of $10\arcmin$.
Fig.~\ref{rm} shows the RM distribution (n=0) for the eastern
filamentary region. We have excluded pixels with a polarized intensity below 
1.40~mK~T$_B$ at $\lambda$11\ cm or 0.75~mK~T$_B$ at $\lambda$6\ cm.

In principle, we require polarization maps at a minimum of three frequencies to
determine the unambiguous RMs. However, we cannot use the $\lambda$21\ cm
map for this purpose because the large-scale diffuse structure dominates
and the filaments are  not clearly visible. RMs
calculated at $\lambda$6\ cm and $\lambda$11\ cm have an ambiguity of
$\rm \pm n\times 362~$rad~m$^{-2}$ (n=0, $\pm$1, $\pm$2, ...). Any value of n other than zero 
seems unlikely
toward the Galactic anti-center because the direction of the local
magnetic field is almost perpendicular to the line of sight and thus
small RM values are expected.

The average RM in the east filamentary region is found to be $-70\pm 8$\ rad
m$^{-2}$, which is almost identical to the RM =$-75\pm 4$\ rad m$^{-2}$ of
an extragalactic source, 0539+266B, shining through this region
\citep{k+88}. Outside the filamentary shell, the RM excess caused by S147 is
likely rather small, as indicated by the smooth distribution of polarized
intensity and polarization angles in Fig.~\ref{11abs}. As was mentioned previously
the average Faraday rotation over the entire area is low. Therefore, the intrinsic 
magnetic field vectors of the
eastern filaments are very close to the magnetic field direction shown
in Fig.~\ref{6cm}.

\subsection{Magnetic fields of the eastern filament}

We estimate the magnetic field strength
in the eastern filamentary region by assuming energy equipartition between
the magnetic field and electrons and protons \citep{fr+04}:
\begin{equation}
B_{\min}  \approx 10 \cdot \Phi \cdot R(deg)\, \mbox{}^{-6/7}
\cdot d(kpc)\, \mbox{}^{-2/7} \cdot S_{1GHz}(Jy)\, \mbox{}^{2/7}
\end{equation}
We integrated the filamentary region and obtained a flux density at
$\lambda$6\ cm of 5.1\ Jy. The flux density at 1\ GHz is extrapolated using
the spectral index of $\alpha$ = $-0.35$. The radius of the radiating region
is assumed to be $0\fdg5$ and the fraction of the radiation volume $\Phi$
is $\sim$ 1. For distances of 0.8\ kpc, we obtain  an estimate for the
magnetic field strength of about 36$\mu$G.
Using the RM value $\rm -70\ rad~m^{-2}$ for the eastern
filamentary region, we estimated the line of sight component of the
intrinsic magnetic field. Assuming an electron density of
100~cm$^{-3}$ within the filaments and their thickness to be $1\arcsec
\sim$ 0.07\ pc \citep{f+85}, we calculate $B_{||}\approx 35\times$
RM/$n_{e}\sim 24 \mu G$. Both values are of the same order. 
Magnetic fields in these filamentary structures are typically 30$\mu$G.

\section{Summary}

We present new sensitive maps of S147 for the total  and
polarized intensity at $\lambda$6\ cm and $\lambda$11\ cm. The $\lambda$6\ cm
maps are the first complete ones obtained so far. Our new measurements
confirm the turnover of the integrated spectrum of S147 at a frequency of
about 1.5~GHz.  The measured structures have been decomposed into
filamentary structures (smaller than $18\arcmin$) and large-scale diffuse
emission component. The T-T plots between 2639\ MHz and 4800\ MHz for these two
emission components prove that the turnover is attributed entirely to
the diffuse emission component, which has a spectral index of $\alpha$
$\sim$-1.35. The filamentary emission component has a spectral index of $\alpha$
$\sim$-0.35 and can be traced up to 40.7\ GHz on WMAP maps.

We discussed several possible mechanisms to explain the turnover
of the diffuse emission component. We can exclude the scenarios of two populations 
of electrons as well as the diffuse shock accelerated of electrons. However,
a compressed magnetic field is a possible explanation. The SNR shock wave
compresses the local magnetic field and shifts the turnover in the Galactic
radio spectrum from about 400\ MHz to about 1.5\ GHz.  High
synchrotron losses, during the early phase of the SNR, would cause a bend at a
rather high frequency, but subsequent expansion of S147 would shift the
frequency toward about 1.5\ GHz as well.

S147 is a faint source located on the near side of the Perseus arm.
Polarized emission from S147 is confused  with large-scale
Galactic emission, except for the bright
filamentary regions. The RM in the eastern filamentary region is about
$-70$\ rad m$^{-2}$. The magnetic field in the filamentary region has
a strength of typically $\sim~30\mu$G.

\begin{acknowledgements}
We acknowledge travel support from the MPG and CAS exchange program.
The $\lambda$6\ cm observations of S147 were carried out at the NanShan
station of Urumqi observatory of NAOC. We thank all the
people involved in the installation of the $\lambda$6\ cm receiver and the
required observing telescope software, in particular, Mr. Otmar Lochner
for the construction of the receiver, and Dr. Peter M\"uller and Dr. Xiaohui Sun
for software installation and development at the telescope. LX would like
to thank Dr. Xiaohui Sun for qualified support on all kinds of data reduction
related issues. We thank Mr. M. Z. Chen, Mr. J. Ma, and the staff of the
Urumqi Observatory for maintenance and assistance during the observations.
The $\lambda$11\
data are based on observations with the 100m telescope of the
Max-Planck-Institut f\"ur Radioastronomie at Effelsberg. The research work
of LX and JLH are supported by the National Natural Science Foundation of
China (10473015 and 10521001) and by the Partner group of MPIfR at
NAOC. We would like to thank Richard Wielebinski and Patrica Reich for critical
reading of the manuscript.
\end{acknowledgements}

\bibliographystyle{aa}
\end{document}